\documentclass[aps,prc,twocolumn,groupedaddress,floatfix,%
               longbibliography]{revtex4-1}
\pdfoutput=1
\usepackage{amsmath,amssymb,latexsym,array,bm}
\usepackage{graphicx}
\usepackage[pdftex,
            pdfauthor={Carl R. Brune},
            pdftitle={},
            pdfsubject={},
            pdfkeywords={},
            pdfproducer={revtex4-1, pdflatex with hyperref},
            pdfcreator={revtex4-1, pdflatex with hyperref}]{hyperref}
\DeclareMathOperator{\Tr}{Tr}

\begin{document}
\title{Secondary $\gamma$-ray decays from the partial-wave \textit{T}~matrix
  with an \textit{R}-matrix application to
  ${}^{15}{\rm N}(p,\alpha_1\gamma){}^{12}{\rm C}$ }
\author{ Carl~R.~Brune }
\email{Electronic address: brune@ohio.edu}
\affiliation{Edwards Accelerator Laboratory,
  Department of Physics and Astronomy,
  Ohio University, Athens, Ohio 45701, USA}

\author{R.~James~deBoer}
\email{Electronic address: rdeboer1@nd.edu}
\affiliation{The Joint Institute for Nuclear Astrophysics}
\affiliation{Department of Physics, University of Notre Dame, Notre Dame,
  Indiana 46556, USA}

\date{\today}

\begin{abstract}
The secondary $\gamma$~rays emitted following a nuclear reaction are
often relatively straightforward to detect experimentally.
Despite the large volume of such data, a practical formalism for describing
these $\gamma$~rays in terms of partial-wave $T$-matrix elements has
never been given. The partial-wave formalism is applicable
when $R$-matrix methods are used to describe the reaction in question.
This paper supplies the needed framework, and it is demonstrated by the
application to the ${}^{15}{\rm N}(p,\alpha_1\gamma){}^{12}{\rm C}$ reaction.
\end{abstract}

\maketitle

\section{Introduction}

We consider here a nuclear reaction sequence
\begin{equation} \nonumber
A(a,b)B \quad \mbox{followed by} \quad B\rightarrow C+\gamma,
\end{equation}
where there are two nuclei in the initial (a and A) and final (b and B)
states and the residual nucleus $B$ is left in an excited state.
Subsequent to the reaction, we have $B\rightarrow C$ by the emission of
a single photon, which we refer to as a secondary $\gamma$~ray.
Note also that $B$ and $C$ correspond to different states in
the {\em same} nucleus.
The purpose of this paper is to describe how the emission of secondary
$\gamma$~rays is correlated with the incident beam direction and/or
the direction of the particle $b$.
We are particularly interested in situations where the transition
matrix for the reaction process is available in partial wave form.

The $R$-matrix theory of nuclear reactions~\cite{Lan58} can in principle
provide a phenomenological description of any reaction process that
only involves two-body channels.
It provides the energy dependence of the partial-wave
transition matrix in terms of resonance parameters.
The use of $R$-matrix methods as a phenomenological tool for analyzing
nuclear reaction data for nuclear astrophysics and other nuclear applications
has become routine~\cite{Des10,Azu11}.
Despite the large volume of data involving secondary $\gamma$~rays,
a practical formalism for calculating secondary $\gamma$-ray emission
has never been given and no currently available $R$-matrix code is capable
of directly analyzing such data.
This paper provides the necessary formalism to make this
type of analysis possible.

From an experimental point of view, the detection of secondary $\gamma$~rays
offers many advantages compared to detection of the outgoing nuclei.
For the detection of these $\gamma$~rays, the experimental energy resolution
is only degraded by Doppler broadening and the intrinsic resolution of
the detector.
Outgoing nuclei are subject to many more types of kinematic broadening,
such as energy loss effects in the case of charged particles,
variation of the kinematics over the acceptance of the detector,
and time-resolution effects in the case of neutron time-of-flight experiments.
A secondary $\gamma$-ray measurement can use a thick target and detectors
which subtend large solid angles without compromising resolution,
leading to very efficient measurements.
Examples where these experimental techniques have been applied include
$^{12}{\rm C}(n,n_1\gamma){}^{12}{\rm C}$~\cite{Nel94},
${}^{16}{\rm O}(n,x\gamma)$~\cite{Nel01}, and
${}^{15}{\rm N}(p,\alpha_1\gamma){}^{12}{\rm C}$~\cite{Bra77}.

However, we do not limit our consideration to cases where the outgoing
nuclei are not detected, as the outgoing nuclei can provide considerable
additional information.
One experimental approach of this type is one in which information about
the direction of the outgoing nuclei is inferred from the
observed Doppler shift of the $\gamma$~ray.
These measurement are very feasible in light nuclei where the $\gamma$~rays
can be detected with $\sim 2$~keV resolution and the range of the
Doppler spread is $\sim 100$~keV.
The exact amount of Doppler spread depends on the kinematics of the
particular reaction and the analysis requires that slowing down effects of
the residual nucleus before it decays are negligible.
When the $\gamma$~rays are detected at $0^\circ$ with respect to the
incident beam direction, this analysis is particularly straightforward.
Examples of experiments of this type are given in
Refs.~\cite{Clo69,Sta70,Try73,Try75,Bra77,Kip99}.

This paper is organized as follows. First, the general formalism for
particle-$\gamma$ correlations is presented. It is then specialized
to the case when the transition matrix is available in partial wave form.
We then further specialize the results to two common experimental
scenarios: the differential cross section for the $\gamma$~ray when
the outgoing nuclei are not detected and the angular distribution of
the reaction products when the $\gamma$~ray is detected at $0^\circ$
relative to the incident beam direction.
Finally, we discuss the implementation of secondary $\gamma$-ray
angular distributions in {\tt AZURE2}~\cite{Azu11} and
show an example application to the
${}^{15}{\rm N}(p,\alpha_1\gamma){}^{12}{\rm C}$ reaction.

\section{General formalism}

The angular distribution of the $\gamma$-ray radiation resulting from the decay
$B\rightarrow C$ is dependent upon the magnetic substate population, or
polarization, of the nucleus $B$ before it decays.
It is thus natural to analyze this information to learn about the
mechanism(s) of such decays.
Important early contributions to the analysis of angular correlations are
provided by \textcite{Sat54} and \textcite{Dev57}.
The formalism used for the calculation of polarization
observables~\cite{Wol56,Wel63,Hei72,Sim74,Sch12}
is also closely related to this topic.

\subsection{Angular correlation function}

A significant paper documenting the understanding of
particle-$\gamma$ angular correlations is the review of angular
correlations in inelastic neutron scattering by \textcite{She63}.
This work is generalized by \textcite{Ryb70} to the case of
arbitrary $b$-$\gamma$ correlations, where
the now standard formalism of \textcite{Ros67}
for describing the $\gamma$~ray is also utilized.
The general approach to angular correlations is summarized by
\textcite[Sec.~10.7]{Sat83} and provides the starting point for our work.
The double differential cross section for detecting $b$ and
the $\gamma$ is given by \cite[Eq.~(10.126)]{Sat83}
\begin{equation}
\frac{{\rm d}^2 \sigma}{{\rm d}\Omega_b{\rm d}\Omega_\gamma} =
\frac{{\rm d}\sigma}{{\rm d}\Omega_b} \frac{W}{4\pi} .
\label{eq:d2sdodo}
\end{equation}
Here, we assume the state $B$ decays via $\gamma$-ray emission
100\% of the time to state $C$.
If this is not the case, the above expression must be multiplied by
the appropriate branching-ratio factor.

This factorization assumes that the process is {\em sequential}, with
the $A(a,b)B$ followed by photon emission. The situation would be much
more complicated if photon emission from the alternative order of
emission, i.e., $A(a,\gamma)[C+b]^*$, was significant.
Such mechanisms could create continuum backgrounds and may also interfere
quantum mechanically with the secondary $\gamma$ ray from $A(a,b)B$.
However, due to the relative weakness of photon emission, such scenarios
are highly unlikely. More quantitatively, photon emission is typically four
or more orders magnitude less likely than nuclear emissions and the
photons would be spread over the energy range allowed phase space.
Such effects could become important in a fine-tuned situation involving
resonances where nuclear emissions are suppressed by isospin or Coulomb /
angular-momentum barrier
considerations and the initial-state photon decay energy matches
that of the secondary photon decay.
These order-of-emission effects are not an issue for any of the
reactions mentioned in this paper.

The angular correlation function $W$ is given by
\cite[Eqs.~(10.127), (10.130), and (10.131)]{Sat83}
\begin{equation}
W=\sum_{kq} t_{kq}(I_B) \, R_k \, \left[\frac{4\pi}{2k+1} \right]^{1/2} Y_{kq}^* ,
\end{equation}
where $t_{kq}(I_B)$ is the polarization tensor of the nucleus $B$,
$R_k$ are the radiation parameters,
and the $Y_{kq}$ are spherical harmonics.
The notation $I_X$ denotes the intrinsic spin of nuclear state $X$.
The $z$ axis is taken to be along the direction of the incident beam.
The polarization tensor  $t_{kq}(I_B)$  is a function of the spherical
coordinates $\Omega_b = (\theta_b,\phi_b)$ defined in the center of mass
of $b+B$.
The spherical harmonic $Y_{kq}$ is a function of the spherical coordinates
$\Omega_\gamma = (\theta_\gamma,\phi_\gamma)$ defined in the
center of mass of nucleus $B$.

\subsection{Radiation parameters}

The $\gamma$ radiation parameters depend upon the intrinsic spins of nuclear
states $B$ and $C$ and the multipolarities present,
and are given in general by
\cite[Eqs.~(10.153) and (10.154)]{Sat83}
\begin{equation}
R_k=\sum_{LL'} g_L \, g_{L'} \, R_k(LL'I_BI_C),
\end{equation}
where
\begin{equation}
\begin{split}
R_k(LL' & I_BI_C) = (2I_B+1)^{1/2}(2L+1)^{1/2} \\
& \times (2L'+1)^{1/2} \times (-1)^{I_B-I_C+L-L'+k+1} \\
& \times (L'L1-1|k0) \, W(LL'I_BI_B;kI_C) .
\end{split}
\label{eq:rk_detail}
\end{equation}
Here, $L$ and $L'$ take on values of the multipolarities of the possible
$\gamma$-ray transitions, \mbox{$(LL'1-1|k0)$} is a Clebsch-Gordan coefficient,
and $W(LL'I_BI_B;kI_C)$ is a Racah coefficient.
The relative multipole amplitudes $g_L$ may be taken to be real and
are normalized such that
\begin{equation}
\sum_L g_L^2=1 ,
\end{equation}
which together with $R_0(LL'I_BI_C)=\delta_{LL'}$ implies $R_0=1$.
Parity considerations require that $k$ only
take on even values, but note Eq.~(\ref{eq:rk_detail}) is not
necessarily zero for $k$ odd~\cite[footnote 16, p.~389]{Sat83}.
Additional properties of the $R_k(LL'I_BI_C)$
are discussed by \textcite{Ros67}.

It is often the case that only one multipole is present, or that
this assumption is a good approximation. In this situation,
there is only a single $L$ value and
\begin{equation}
\begin{split}
R_k = & \, (2I_B+1)^{1/2}(2L+1)(-1)^{I_B-I_C+k+1} \\
  & \times (LL1-1|k0) \, W(LLI_BI_B;kI_C) .
\end{split}
\end{equation}
If two multipoles, $L_1$ and $L_2$, are present, then one may define the
mixing ratio $\delta=g_{L_2}/g_{L_1}$ and~\cite[Eq.~(15)]{Ryb70}
\begin{equation}
\begin{split}
R_k = & \left[ R_k(L_1L_1I_BI_C)+2\delta R_k(L_1L_2I_BI_C) \right. \\
  & \left. + \, \delta^2 R_k(L_2L_2I_BI_C)\right] / (1+\delta^2).
\end{split}
\end{equation}

\subsection{Polarization tensors}

The polarization tensors $t_{kq}(I_B)$ describe the polarization
of the final nucleus $B$; the precise definition is given in terms of
expectation values of tensor operators $\tau_{kq}(I_B)$ in
Ref.~\cite[Secs.~10.3.2 and~10.3.3]{Sat83}.
Specifically, we have \cite[Eq.~10.32]{Sat83}
\begin{equation}
t_{kq}(I_B)=\frac{\Tr[{\bm T}{\bm T}^\dag \tau_{kq}(I_B)]}
{\Tr[{\bm T}{\bm T}^\dag]},
\end{equation}
where
${\bm T}$ is the transition matrix for $A(a,b)B$ and the trace implies
a sum over all spin projections of the incoming and outgoing nuclei.
To make this more explicit, we note~\cite[Eq.~(10.25b)]{Sat83}
\begin{equation}
\begin{split}
\langle I_B m_B|\tau_{kq}(I_B)| & I_B {m_B}'\rangle \\
  & = (2k+1)^{1/2} (I_Bk m_B q|I_B {m_B}')
\end{split}
\end{equation}
and take ${\bm T}$ in the individual particle spin basis to be
${\mathcal M}_{m_bm_B:m_am_A}(\Omega_b)$, which is also known
as the scattering amplitude.
We now have
\begin{equation}
\begin{split}
t_{kq} & (I_B) = (2k+1)^{1/2} \\
  & \times \Biggl[ \,\, \sum_{m_b m_B {m_B}' m_a m_A}
  (I_Bkm_Bq|I_B{m_B}') \\
  & \quad\quad\times
  {\mathcal M}_{m_b{m_B}':m_am_A}(\Omega_b)
  {\mathcal M}_{m_bm_B:m_am_A}^*(\Omega_b) \Biggr] \\
  & \, / \displaystyle \sum_{m_bm_Bm_am_A} |
  {\mathcal M}_{m_bm_B:m_am_A}(\Omega_b)|^2 ,
\end{split}
\end{equation}
where $m_B'=m_B+q$ is required for the Clebsch-Gordan coefficient
to be non-zero.
Also note that $t_{00}=1$ and that the differential cross section for
particle $b$ is given by
\begin{equation} \label{eq:dsdo_b}
\begin{split}
\frac{{\rm d}\sigma}{{\rm d}\Omega_b}= & \, \frac{1}{(2I_A+1)(2I_a+1)} \\
  & \times \sum_{m_bm_Bm_am_A} |{\mathcal M}_{m_bm_B:m_am_A}(\Omega_b)|^2.
\end{split}
\end{equation}

The $\phi_b$ dependence of the scattering amplitude
$\mathcal{M}$ is very simple and allows another scattering amplitude
$f_{m_bm_B:m_am_A}$ that only depends upon $\theta_b$ to be defined:
\begin{equation}
\begin{split}
{\mathcal M}_{m_bm_B:m_am_A}(\Omega_b) = & \, e^{i(m_b+m_B-m_a-m_A)\phi_b} \\
  & \times f_{m_bm_B:m_am_A}(\theta_b).
\end{split}
\end{equation}
This is the form of the scattering amplitude calculated by the
computer code {\sc fresco}~\cite{Tho88}.
It is also easy to show that
\begin{equation}
t_{kq}(\Omega_b) = e^{iq\phi_b} t_{kq}(\theta_b,0),
\end{equation}
as expected for the tensor operator $t_{kq}$.

This form of the scattering amplitude may be sufficient for
some calculations, e.g., if the scattering amplitudes from the
aforementioned {\sc fresco} code are available.
For example, Eq.~(\ref{eq:d2sdodo}) may be integrated over $\Omega_b$ to yield
the differential cross section for $\gamma$-ray emission:
\begin{align}
\frac{d\sigma}{d\Omega_\gamma} &= \sum_k R_k P_k(\cos\theta_\gamma)
  \int_{-1}^1 \frac{d\cos\theta_b}{2} \frac{d\sigma}{d\Omega_b}
  t_{k0}(I_B) \label{eq:diff_gamma_1} \\
\begin{split}
  &= \, \frac{1}{(2I_A+1)(2I_a+1)}\sum_k R_k P_k(\cos\theta_\gamma) \\
  & \quad \times \frac{(2k+1)^{1/2}}{2} \int_{-1}^1 d\cos\theta_b \\
  & \quad \times \sum_{m_bm_Bm_am_A} |f_{m_bm_B:m_am_A}(\theta_b)|^2 \\
  & \quad \times (I_Bkm_B0|I_Bm_B) ,
\end{split}
\end{align}
where $P_k$ are the Legendre polynomials.

\section{Partial-wave \textit{T} matrix}

The $T$~matrix can then in principle be used to calculate any experimental
observable, with the calculation being independent of the model used to
determine the $T$~matrix.
In this section, the $T$~matrix is converted to partial wave form
and specific $\gamma$-ray observables are calculated.
The material in Secs.~\ref{subsec:dsdo_gamma},  \ref{subsec:correl_gen},
and~\ref{subsec:gamma0} represents the main results of this work.
Note that in an $R$-matrix approach, the partial-wave $T$~matrix is
calculated from the $R$-matrix or level matrix.

The scattering amplitudes may be constructed from the partial-wave
$T$~matrix as follows.
According to \textcite[Eq.~VIII.2.3, p.~292]{Lan58},
the scattering amplitudes connecting nonelastic channels
are given in the channel spin basis by
\begin{equation}
\begin{split}
& A_{bB s'\nu':aA s\nu} = i\frac{\pi^{1/2}}{k_{aA}} \sum_{JM\ell\ell'm'}
  (2\ell+1)^{1/2} \\
  & \quad \times (s\ell\nu0|JM) (s'\ell'\nu'm'|JM) \\
  & \quad \times Y_{\ell'm'}(\Omega_b) T^J_{bBs'\ell':aAs\ell} ,
\end{split}
\end{equation}
where $k_{aA}$ is the center-of-mass wave number in the $a+A$ system and
$T^J_{bBs'\ell':aAs\ell}$ is the partial-wave $T$~matrix.
In the individual particle spin basis, the scattering amplitudes become
\begin{equation}
\begin{split}
& \mathcal{M}_{m_bm_B:m_am_A}(\Omega_b) = i\frac{\pi^{1/2}}{k_{aA}} \\
  & \quad \times \sum_{JM\ell\ell'm's\nu s'\nu'}
  (2\ell+1)^{1/2} (s\ell\nu0|JM) \\
  & \quad \times (s'\ell'\nu'm'|JM) (I_aI_Am_am_A|s\nu) \\
  & \quad \times (I_bI_Bm_bm_B|s'\nu')
  Y_{\ell'm'}(\Omega_b) T^J_{bBs'\ell':aAs\ell} ,
\end{split}
\end{equation}
which is suitable for calculating the general particle-$\gamma$ correlation
function.

\subsection{Differential cross section for particle \texorpdfstring{$b$}{b}}

The differential cross section for particle $b$ can then in principle
be calculated from $\mathcal{M}_{m_bm_B:m_am_A}(\Omega_b)$
using Eq.~(\ref{eq:dsdo_b}).
This result can, in a certain sense, be simplified by introducing
two sets of summing indices (one for each occurrence of $\mathcal{M}$),
expressing the products of spherical harmonics as sums of single spherical
harmonics, and contracting the sums over magnetic substates.
The result is
\begin{equation} \label{dsdo_b}
\frac{{\rm d}\sigma}{{\rm d}\Omega_b}= \frac{1}{(2I_A+1)(2I_a+1)} \,
  \frac{\pi}{k_{aA}^2} \sum_\mathcal{K} G_\mathcal{K}(\theta_b),
\end{equation}
where
\begin{equation}
\begin{split}
&G_\mathcal{K}(\theta_b) = \sum_{J_1 J_2 \ell_1 \ell_2 \ell_1' \ell_2' ss'}
  \hspace*{-0.2in} (-1)^{s-s'} (2J_1+1)(2J_2+1) \\
  & \quad \times [(2\ell_1+1)(2\ell_2+1)(2\ell_1'+1)(2\ell_2'+1)]^{1/2} \\
  & \quad \times (\ell_1 \ell_2 00 | \mathcal{K} 0) \,
  (\ell_1' \ell_2' 00 | \mathcal{K} 0) \,
  W(l_1 J_1 l_2 J_2 ;s\mathcal{K}) \\
  & \quad \times W(l_1' J_1 l_2' J_2 ;s'\mathcal{K}) \,
  \frac{P_{\mathcal{K}}(\cos\theta_b)}{4\pi} \\
  & \quad \times T^{J_1 *}_{bBs'\ell_1':aAs\ell_1} T^{J_2}_{bBs'\ell_2':aAs\ell_2} ,
\end{split}
\end{equation}
which agrees with Eq.~VII.2.6 of \textcite[p.~292]{Lan58}.

\subsection{Differential cross section for \texorpdfstring{$\gamma$}{gamma}-ray
  emission}
\label{subsec:dsdo_gamma}

The differential cross section for $\gamma$-ray emission may be found by
integrating Eq.~(\ref{eq:d2sdodo}) over $\Omega_b$:
\begin{equation} \label{eq:dsdo_gamma}
\begin{split}
  \frac{d\sigma}{d\Omega_\gamma} = & \, \frac{1}{(2I_A+1)(2I_a+1)}
  \frac{\pi}{k_{aA}^2} \sum_k (2k+1)^{1/2} \\
  & \times R_k \, \frac{P_k(\cos\theta_\gamma)}{4\pi} \, H_k,
\end{split}
\end{equation}
where
\begin{equation}
\begin{split}
  & H_k = \frac{k_{aA}^2}{\pi} \int_{4\pi} d\Omega_b \\
  & \quad \times \sum_{m_bm_Bm_am_A}
  |\mathcal{M}_{m_bm_B:m_am_A}(\Omega_b)|^2 \\
  & \quad \times (I_Bkm_B0|I_Bm_B) .
\end{split}
\end{equation}
This equation may be ``simplified'' in a similar manner, except that the
integration over $\Omega_b$ is carried out using the orthogonality of
the spherical harmonics.
This procedure results in
\begin{equation} \label{eq:Hk_1}
\begin{split}
& H_k = \sum_{J_1J_2\ell_1\ell_2\ell'ss_1's_2'} \hspace*{-0.2in}
  (-1)^{k+s_2'-s_1'} (2J_1+1)(2J_2+1) \\
  & \quad \times [(2\ell_1+1)(2I_B+1)(2s_1'+1)(2s_2'+1)]^{1/2} \\
  & \quad \times (k\ell_100|\ell_20) \,
  W(kI_Bs_2'I_b ; I_Bs_1') \\
  & \quad \times W(ks_1'J_2\ell' ; s_2'J_1)
  W(k J_1 \ell_2 s ; J_2 \ell_1) \\
  & \quad \times T^{J_1 *}_{bBs_1'\ell':aAs\ell_1} T^{J_2}_{bBs_2'\ell':aAs\ell_2} .
\end{split}
\end{equation}
Note that the factor of $(-1)^k$ in this expression does not actually come
into play since $k$ is required to be even. An equivalent expression with
the arguments of the angular momentum coupling functions arranged in a
more symmetrical and standard manner is
\begin{equation}
\begin{split}
& H_k = \sum_{J_1J_2\ell_1\ell_2\ell'ss_1's_2'}
  (-1)^{\ell' + s - 2s_1' - I_B + I_b - J_1 -J _2 } \\
  & \quad \times (2J_1+1)(2J_2+1)(2\ell_1+1)^{1/2} \\
  & \quad \times \left[\frac{(2\ell_2+1)(2I_B+1)(2s_1'+1)(2s_2'+1)}
      {2k+1}\right]^{1/2} \\
  & \quad \times (\ell_1 \ell_2 00|k0) \,
  W(I_B s_1' I_B s_2' ; I_b k) \\
  & \quad \times W(s_1' J_1 s_2' J_2 ; \ell' k)
  W(\ell_1 J_1 \ell_2 J_2 ; s k) \\
  & \quad \times T^{J_1 *}_{bBs_1'\ell':aAs\ell_1} T^{J_2}_{bBs_2'\ell':aAs\ell_2} .
\end{split}
\end{equation}

\subsection{General correlation function}
\label{subsec:correl_gen}

The general correlation function corresponding to Eq.~(\ref{eq:d2sdodo})
can be written
\begin{equation} \label{eq:correl_gen}
\begin{split}
  & \frac{{\rm d}^2 \sigma}{{\rm d}\Omega_b{\rm d}\Omega_\gamma} =
  \frac{1}{(2I_A+1)(2I_a+1)} \frac{\pi}{k_{aA}^2} \\
  & \quad \times \sum_{kq} R_k \, \frac{Y_{kq}^*(\Omega_\gamma)}{(4\pi)^{1/2}}
  \, F_{kq}(\Omega_b),
\end{split}
\end{equation}
where
\begin{equation}
\begin{split}
& F_{kq}(\Omega_b) = \frac{k_{aA}^2}{\pi} \sum_{m_b m_B {m_B}' m_a m_A}
  (I_Bkm_Bq|I_B{m_B}') \\
  & \quad \times \mathcal{M}_{m_bm_B:m_am_A}^*(\Omega_b)
  \mathcal{M}_{m_b{m_B}':m_am_A}(\Omega_b) .
\end{split}
\end{equation}
The contraction of the magnetic substate quantum numbers is similar to that
described above.
This calculation is also closely related to the results derived in
Refs.~\cite{Wel63,Hei72}, but we do not rotate the $t_{kq}$ to make
the $z$~axis along the direction of the scattered particle.
We find
\begin{equation} \label{eq:fkq}
\begin{split}
F_{kq}&(\Omega_b) = \sum_{J_1 J_2 \ell_1 \ell_2 \ell_1' \ell_2' s s_1'
    s_2'\mathcal{K} \mathcal{L}}
  \hspace*{-0.25in}(-1)^{\ell_2'+s+s_2'-I_B+I_b+J_2} \\
  & \times (2J_1+1)(2J_2+1) \\
  & \times \left[\frac{(2\ell_1+1)(2\ell_2+1)(2I_B+1)(2\mathcal{K}+1)}%
    {2\mathcal{L}+1}\right]^{1/2} \\
  & \times \left[(2\ell_1'+1)(2\ell_2'+1)
    (2s_1'+1)(2s_2'+1)\right]^{1/2} \\
  & \times  (\ell_1 \ell_2 00|\mathcal{K}0) \,
  (\ell_1' \ell_2' 00|\mathcal{L}0) \, (\mathcal{K}k0q|\mathcal{L}q) \\
  & \times  W(I_B s_1' I_B s_2' ; I_b k) \,
    W(\ell_1 J_1 \ell_2 J_2 ; s\mathcal{K}) \\
  & \times \left\{ \begin{array}{ccc}
      \ell_1' & s_1' & J_1 \\
      \mathcal{L} & k & \mathcal{K} \\
      \ell_2' & s_2' & J_2 \end{array} \right\} \,
  \frac{Y_{\mathcal{L}q}^*(\Omega_b)}{(4\pi)^{1/2}} \\
  & \times T^{J_1 *}_{bBs_1'\ell_1':aAs\ell_1} T^{J_2}_{bBs_2'\ell_2':aAs\ell_2} ,
\end{split}
\end{equation}
where $\{\}$ denotes the 9-$J$ symbol. The previous results can be
expressed as specializations of this formula:
\begin{equation}
\sum_\mathcal{K} G_\mathcal{K}(\theta_b) = F_{00}(\Omega_b)
\end{equation}
and
\begin{equation}
H_k = \int_{4\pi} d\Omega_b F_{k0}(\Omega_b) .
\end{equation}

\subsection{Correlation function for
  \texorpdfstring{$\theta_\gamma=0$}{theta(gamma)=0}}
\label{subsec:gamma0}

When the $\gamma$~rays are detected at $0^\circ$ with respect to the
incident beam direction, their Doppler shift only depends on the
angle $\theta_b$ of the ejectile. In this situation, it is straightforward
to analyze the $\gamma$-ray energy spectrum to deduce the experimental
correlation function~\cite{Clo69,Sta70,Try73,Try75,Bra77,Kip99}.
When $\theta_\gamma=0$, Eq.~(\ref{eq:correl_gen}) becomes
\begin{equation}
\begin{split}
  & \frac{{\rm d}^2 \sigma}{{\rm d}\Omega_b{\rm d}\Omega_\gamma}
  (\theta_\gamma=0) =
  \frac{1}{(2I_A+1)(2I_a+1)} \frac{\pi}{k_{aA}^2} \\
  & \quad \times \sum_{kq} R_k \, \left(\frac{2k+1}{4\pi}\right)^{1//2}
  \, F_{k0}(\theta_b) .
\end{split}
\end{equation}
The quantity $F_{k0}(\theta_b)$ is calculated using Eq.~(\ref{eq:fkq})
with $q=0$, where there is little simplification, except for the
replacement
\begin{equation}
  \frac{Y_{\mathcal{L}q}^*(\Omega_b)}{(4\pi)^{1/2}} \rightarrow
  \frac{(2\mathcal{L}+1)^{1/2}}{4\pi} P_\mathcal{L}(\cos\theta_b) .
\end{equation}

\section{Implementation in {\tt AZURE2}}

The capability of fitting the differential cross section for $\gamma$-ray
emission has been implemented in the computer code {\tt AZURE2}~\cite{Azu11}.
This observable is calculated from the partial-wave $T$~matrix
using Eqs.~(\ref{eq:dsdo_gamma}) and (\ref{eq:Hk_1}).
This calculation is very similar in structure to the calculation
of the differential cross section for the emission of particle $b$
given by Eq.~(\ref{dsdo_b}), an observable that is already
calculated in {\tt AZURE2}. This modification thus proved to be
a relatively simple addition to the computer code.
An example is presented in the following section.

\section{Application to
  \texorpdfstring{${}^{15}{\rm N}(p,\alpha_1\gamma){}^{12}{\rm C}^*$}%
  {15N(p,alpha1)12C*}}

\begin{figure}[tb]
\includegraphics[width=\columnwidth]{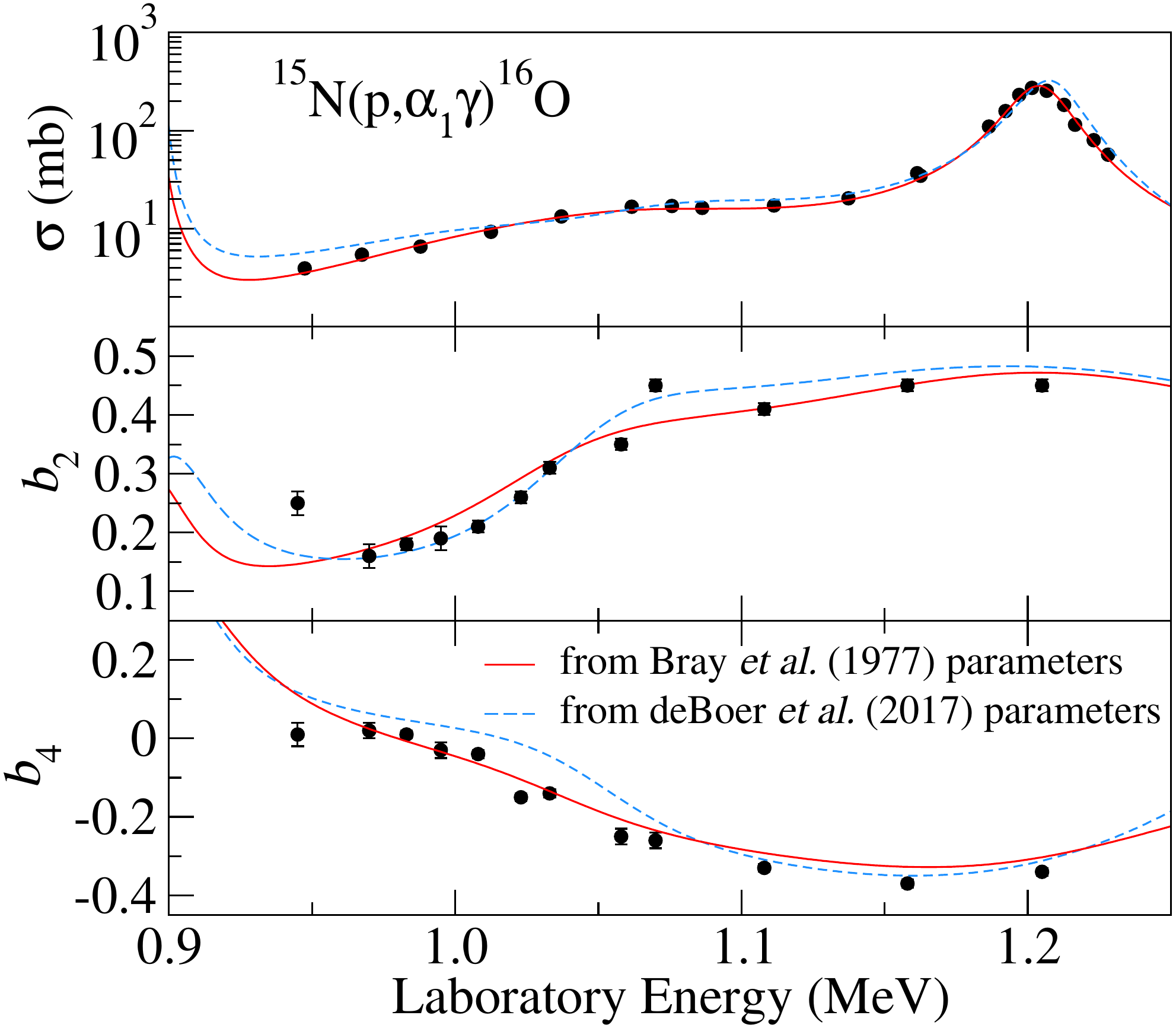}
\caption{(Color online) The total cross section and $\gamma$-ray angular
distribution coefficients for the
${}^{15}{\rm N}(p,\alpha_1\gamma){}^{12}{\rm C}^*$ reaction.
The experimental results of \textcite{Bra77} are shown as points.
Calculations using the \textcite{Bra77} and \textcite{deB17} $R$-matrix
parameters are shown as the solid red and dashed blue curves,
respectively.} \label{fig:bray}
\end{figure}

Measurements of the ${}^{15}{\rm N}(p,\alpha_1\gamma){}^{12}{\rm C}^*$
reaction in the vicinity of 1-MeV proton energy have been
performed by \textcite{Bra77}.
Data for the total cross section and $\gamma$-ray angular
distribution coefficients are given in Fig.~2 and Table~1 of their work.
These coefficients are defined by
\begin{equation} \label{eq:dsdo_bk}
\frac{d\sigma}{d\Omega_\gamma} = \frac{\sigma_{\rm tot}}{4\pi}
\sum_{k} b_k P_k(\cos\theta_\gamma),
\end{equation}
where $\sigma_{\rm tot}$ is the total cross section for populating the
first excited state of ${}^{12}{\rm C}$ and $b_0\equiv 1$.
The $\gamma$-ray transition from the first excited state to the ground state
of ${}^{12}{\rm C}$ is $2^+\rightarrow 0^+$ and is a pure $E2$ transition.
Equations (\ref{eq:dsdo_gamma}) and~(\ref{eq:dsdo_bk}) thus only
receive contributions from $k=0$, 2, and~4.
The total cross section and coefficients $b_2$ and $b_4$ measured
by \textcite{Bra77} are shown by the points in Fig.~\ref{fig:bray}.

\textcite{Bra77} also performed an $R$-matrix fit to their
${}^{15}{\rm N}(p,\alpha_1\gamma){}^{12}{\rm C}^*$ measurements and
other ${}^{15}{\rm N}+p$ reaction data.
They included the $b_k$ coefficients in their fit, but unfortunately
no formulas or other descriptions of their procedures were given.
They do , however, provide in Tables~2 and~3 their best fit $R$-matrix
parameters, in the Lane and Thomas formalism~\cite{Lan58}.
Using these parameters in {\tt AZURE2}, we have calculated
$\sigma_{\rm tot}$, $b_2$, and $b_4$, with the results being very close
to the calculations given by \textcite{Bra77} in Fig.~2 of their paper.
Our calculations are also shown as the solid curve in our Fig.~\ref{fig:bray},
where they are seen to be very close to the experimental results
of \textcite{Bra77}.

A comprehensive $R$-matrix analysis of the ${}^{16}{\rm O}$ compound
nucleus was recently published by \textcite{deB17}.
The angular distribution coefficients $b_2$ and $b_4$ defined above
were not included in the fit or otherwise considered in the analysis.
However, the $R$-matrix parameters given by Ref.~\cite{deB17} provide 
an excellent description of these coefficients.
In this case, the $R$-matrix parameters are given in the alternative
$R$-matrix formalism~\cite{Bru02}.
The $\sigma_{\rm tot}$, $b_2$, and $b_4$ resulting from using these
parameters in an {\tt AZURE2} calculation are shown as the dashed curves
in Fig.~\ref{fig:bray}.
These calculations are seen to be very close to both the \textcite{Bra77}
data and the calculations using their parameters.
This example demonstrates that the $R$-matrix formalism can be used
to predict the $\gamma$-ray angular distributions, provided the parameters
can be suitably constrained using other reaction data.

\textcite{Bra77} also presented $\gamma$-ray energy spectra measured
at $\theta_\gamma=0$ that were also well described by their $R$-matrix fit.
Again, no formulas or other descriptions of their procedures were given.
Their data are shown in their Fig.~3, and unfortunately are not
presented in a manner that facilitates a simple comparison with
other calculations of the correlation function for $\theta_\gamma=0$.

\section{Conclusions}

The general formalism for calculating the correlation of secondary
$\gamma$ rays with the beam direction and reaction products has been reviewed.
The general results have been adapted to the partial-wave $T$-matrix formalism
and specializations to common special cases have been presented.
These specializations are well suited for calculations when the
the reaction of interest is described by $R$-matrix theory.
The case of the $\gamma$-ray angular distribution, with the other
reaction products undetected, has been implemented in the $R$-matrix
code {\tt AZURE2}. This capability has been demonstrated with
the example of the ${}^{15}{\rm N}(p,\alpha_1\gamma){}^{12}{\rm C}^*$
reaction. We anticipate the application of this type of analysis
to several other cases. Additional measurements have already been performed
at the University  of Notre Dame of the
${}^{15}{\rm N}(p,\alpha_1\gamma){}^{12}{\rm C}^*$
reaction, extending the range of study to higher energies~\cite{deB20}. 

At last two extensions to this work can be envisioned.
One additional possibility is the situation where the initial reaction
is a radiative capture reaction.
Another is the case of $\gamma$-ray cascades following
a reaction. This scenario can be addressed using the formalism
given by \textcite{Ros67}.

\begin{acknowledgments}
This work was supported in part by the U.S. Department of Energy,
under Grants No.~DE-NA0003883 and No.~DE-FG02-88ER40387 at Ohio University,
and the National Science Foundation under Grants No.~PHY-1713857 (JINA-CEE)
and No.~PHY-1430152 at the University of Notre Dame.
\end{acknowledgments}

\bibliography{angcor.bib}

\end{document}